\documentstyle[11pt,newpasp,psfig,subfigure,twoside]{article}
\index{Parkes}
\index{pulsars!millisecond}
\index{pulsars!surveys}
\index{radio}

\begin{document}

\title{Searching for sub-millisecond pulsars from highly polarized 
radio sources}

\author{J.~L. Han$^1$, R.~N. Manchester$^2$, A.~G. Lyne$^3$ and
 G. J. Qiao$^4$
        }
\affil{$^1$ National Astronomical Observatories, CAS, Beijing 100012, China\\
$^2$ Australia Telescope National Facility, CSIRO, Epping,
         Australia\\
$^3$ University of Manchester, Jodrell Bank Observatory, SK11 9DL,
         UK\\
$^4$ Department of Astronomy, Peking University,
        Beijing 100871, China\\
	} 

\begin{abstract}
Pulsars are among the most highly polarized sources in the universe.
The NVSS has catalogued 2 million radio sources with linear
polarization measurements, from which we have selected 253 sources,
with polarization percentage greater than 25\%, as targets for pulsar
searches. We believe that such a sample is not biased by selection
effects against ultra-short spin or orbit periods. Using the Parkes
64m telescope, we conducted searches with sample intervals of 0.05 ms and
0.08 ms, sensitive to submillisecond pulsars. Unfortunately we did not
find any new pulsars.
\end{abstract}

\vspace{-3mm}
\section{Introduction}

The NRAO VLA Sky Survey (NVSS; Condon et al. 1998) contains 1.8 million
radio sources with flux density at 1400 MHz greater than 2.5 mJy with linear
polarization information.  Many pulsars are known to have high linear
polarization on average. Han \& Tain (1999) identified 97 known pulsars from
the NVSS source catalogue and they found that, on average, the linear
polarization percentage of pulsars is much higher than that of other classes
of objects such as quasars and BL-Lac objects. Therefore, high linear
polarization can be used as a criterion for selecting pulsars with all
kinds of possible periods, including submillisecond pulsars for which
normal untargetted surveys are presently impossible.
Submillisecond pulsars could be strange-quark stars, which have been explored 
theoretically (e.g. Madsen 1998), but with no detection observationally
yet (e.g. Edwards et al. 2001).

We have selected 253 unresolved sources with high linear polarization
(linear polarization percentage $L/S > 25\%$; the uncertainty of $L/S <
10\%$) from the NVSS catalog and searched for pulsed emission. These sources
have a flux density which is generally larger than $\sim$4~mJy.
To have been missed by previous
pulsar surveys, they must have some combination of the following properties:
short (millisecond) pulse period, short orbital period, and/or high
dispersion. Most of these sources are at relatively high Galactic 
latitudes and so the first two properties are likely to be the most
important if they are pulsars.

\begin{figure}[thb]
\psfig{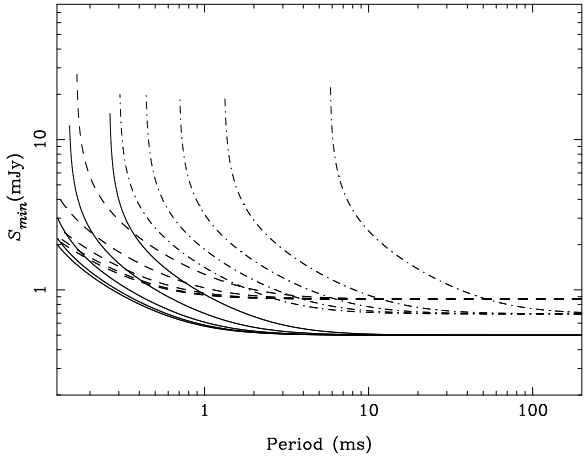}
\caption{The minimum detectable flux density at 8$\sigma$,
assuming the pulse has
a width of 10\% of period. Solid lines are for the 512x0.5MHz system, and 
dashed lines for the 256x0.25MHz system, each with 5 DM values (bottom to top):
0, 20, 40, 80, and 160 pc~cm$^{-3}$. Dash-dotted lines are for the
converted sensitivity of Parkes 70~cm pulsar survey. Our searches 
are sensitive to sub-millisecond pulsars.}
 \end{figure}

We searched for short-period pulsed emission from these selected
sources by using Parkes 64m telescope in October 8-19 1999 and
December 9-11 1999. The central beam of the 20-cm multibeam system was
used, having a system temperature of 21~K and a gain of 0.735
K~Jy$^{-1}$. The two orthogonally polarized signals were amplified and
fed to a filter-bank system, which was used in one of two
configurations: 512 channels of 0.5~MHz bandwidth centered at 1261.75
MHz, or 256 channels of 0.25~MHz bandwidth centered on 1293 MHz. After
detection of each channel, the two polarization signals were added
together and 1-bit digitized. Data were sampled every 0.08~ms for a
total observation time of 340 seconds for the 512x0.5~MHz system, and
0.05~ms sample time for 450 seconds for the 256x0.25~MHz system.

Data-processing consisted of sub-band dedispersion, fine-dedispersion
and periodicity search. As shown in Fig.1, our search is very
sensitive to submillisecond pulsars up to a dispersion measure of 80
pc~cm$^{-3}$, much better than the Parkes 70~cm pulsar survey
(Manchester et al. 1996) which had a sensitivity of 3~mJy at 400 MHz
for long period pulsars, corresponding to about 0.5 mJy at 1400 MHz
for a typical pulsar spectral index of $-1.5$. That survey was
insensitive to pulsars with period less than 2 ms for typical DMs.

No new pulsars were detected though test observations of known pulsars
gave results as expected.  This null result implies that either radio
pulsars do not have pulse periods of less than 1 ms, or submillisecond
pulsars are not highly polarized radio sources.

\end{document}